\documentclass[%
preprint, 
aps, 
]{revtex4-2}

\usepackage{amsmath}
\usepackage{amssymb}
\usepackage{graphicx}

\usepackage{setspace}
\usepackage{ragged2e} 
\usepackage{dcolumn}
\usepackage{soul}
\usepackage{comment}
\usepackage{color}
\usepackage{multirow}
\usepackage{ulem}

\newcommand{\blue}[1]{\textcolor{blue}{#1}}

\usepackage[mathlines]{lineno}
\usepackage{hyperref}

\begin{document}

\title{Marangoni swimmer pushing particle raft under 1D confinement}

\author{ {\large Abhradeep Maitra$^1$, Anupam Pandey$^2$, Sebastien Michelin$^3$, Sunghwan Jung$^4$} }

\address{ {\footnotesize \mbox{$^1$ Department of Mechanical and Aerospace Engineering, Cornell University, Ithaca, NY 14853, USA }\\ \mbox{$^2$ Mechanical and Aerospace Engineering, Syracuse University, Syracuse, USA}\\ \mbox{$^3$ LadHyX – CNRS, Ecole Polytechnique -- Institut Polytechnique de Paris, Palaiseau, France} \\ \mbox{$^4$ Department of Biological and Environmental Engineering, Cornell University, Ithaca, NY 14853, USA }\\}}
    
\email{Correspondence to: sj737@cornell.edu, am3326@cornell.edu}

\keywords{}

\begin{abstract}
Active matter systems, due to their spontaneous self-propulsion ability, hold potential for future applications in healthcare and environmental sustainability. Marangoni swimmers, a type of synthetic active matter, are a common model system for understanding the underlying physics. Existing studies of the interactions of active matter with passive particles have mostly focused on the modification of the  behavior of the passive particles. In contrast, we analyse here experimentally the impact on the self-propulsion of camphor-infused agarose disks (active) of their interactions with floating hollow glass microspheres (passive) within an annular channel. Two distinct regimes are observed: a steady regime with uni-directional motion of the swimmer at low packing fractions ($\phi_{\textrm{ini}} \lesssim 0.45$) and an oscillatory regime with to-and-fro motion at higher packing fractions ($\phi_{\textrm{ini}} \gtrsim 0.45$). In the former, the swimmer pushes nearly the entire particle raft together with it, like a towing cargo, causing a decrease in swimmer speed with increasing packing fraction due to the additional drag from the particle raft. A simplified force-balance model is finally proposed that captures the experimental trend in swimmer speed reasonably well.
\end{abstract}

\maketitle

\section{Introduction} \label{Introduction}

Active matter comprises of systems which can spontaneously consume energy, provided by internal sources or their environment, in order to generate motion \cite{marchetti2013hydrodynamics, bechinger2016active}. The intense research interest in recent years for such systems has been motivated by their intriguing collective and emergent behaviors such as self-organization \cite{halloy2007social,riedel2005self} instabilities\cite{buttinoni2013dynamical,bechinger2016active,saintillan2007orientational}, clustering \cite{palacci2013living,buttinoni2013dynamical} and swarming \cite{vicsek1995novel}, commonly observed in biological active systems such as bird flocks, fish schools, insect swarms etc \cite{bechinger2016active,michelin2023self}. Moreover, micro- and nano- synthetic active systems offer potential applications in targeted drug delivery \cite{wang2012nano,nelson2010microrobots,patra2013intelligent} and environmental remediation\cite{gao2014environmental,jurado2018micromotors}.\\

To investigate the rich physics of active matter systems, researchers
have developed and studied a variety of synthetic active swimmers. Typically, these swimmers are either externally actuated, using electric fields (e.g. Quincke rollers \cite{bricard2013emergence}), thermal gradients \cite{palacci2013living} or magnetic fields \cite{ghosh2009controlled}, or they are self-propelled by releasing internally stored chemicals. Further, these chemically-active swimmers can be categorized into two types : one with intrinsic asymmetry, such as Janus particles \cite{dreyfus2005microscopic,moran2017phoretic} which have non-uniform chemical distribution on their surface, and the other with intrinsic symmetry in terms of the chemical distribution as well as geometry \cite{nagayama2004theoretical,izri2014self,michelin2023self}. The latter type self propels by spontaneous symmetry breaking emerging from the non-linear coupling between the local chemical field and fluid flow \cite{michelin2013spontaneous,michelin2023self}. Among these, isotropic Marangoni swimmers, focus of this study, are a classical system \cite{tomlinson1862ii,nakata1997self}.\\

Isotropic Marangoni swimmers trapped at a free surface self-propel by releasing tensio-active molecules on the interface, and spontaneously breaking the symmetry of surface tension around them through the convective transport of these surfactants by the Marangoni flows they generate \cite{suematsu2014quantitative}. Several studies have explored the dynamics of individual swimmers in different one-dimensional channel geometries such as linear channel \cite{koyano2016oscillatory,hayashima2001camphor} and annular channel \cite{nagayama2004theoretical,shimokawa2018power,nakata2013motion}. In annular channel, swimmers were observed to perform unidirectional motion while introducing a partition into the channel altered it to back-and-forth motion \cite{nagayama2004theoretical}. Along with experiments, analytical \cite{boniface2019self,suematsu2014quantitative,boniface2021role}  as well as numerical models for reproducing the swimmer motion and linear stability analysis for explaining the bifurcation in the dynamics have been presented. Further, experiments in dumbbell shaped channels \cite{nakata2000spontaneous, upadhyaya2024stochastic} revealed a mode-switching behavior. More recently, experiments in a circular domain showed rich dynamics dominated by motion along the boundary and circular motion in the bulk\cite{cruz2024active}. The motion in the bulk was also shown to be consistent with active chiral particle dynamics. Furthermore, `run', `tumble' and `rest' like states were shown to exist for a swimmer in a flower-shaped container \cite{dey2025run}. There were also studies on exploring the interaction of multiple swimmers and their emergent behaviors \cite{kohira2001synchronized, mottammal2021collective, bourgoin2020kolmogorovian,gouiller2021mixing,gouiller2021two,ikura2013collective,nakata2015physicochemical}. \\
 
 Lately, there has been a focus on interaction of active swimmers with passive particles in its environment, aiming to mimic more realistic conditions\cite{bechinger2016active}. Studies with different ratio of active to passive particles have been performed. Dynamic clustering and surface melting behaviors have been shown for a bath of colloidal particles with a small concentration of active swimmers \cite{kummel2015formation}. Partial mixing and turbulent-like spectra was observed for interfacial floaters in presence of few Marangoni swimmers \cite{gouiller2021mixing}.  Some numerical studies for systems with equal amount of active and passive particles reported spontaneous segregation \cite{mccandlish2012spontaneous} and phase separation like phenomena \cite{yang2012using,takatori2015theory,farage2015effective,tung2016micro}. Motion of passive particles in a bath of active swimmers (bacteria) was shown to be super-diffusive at short timescales due to the collective dynamics of the bacteria \cite{wu2000particle}. Moreover, even the long time diffusivity of passive particles was enhanced in a suspension of active swimmers and it showed a linear increasing trend with activity \cite{mino2011enhanced}. Majority of the focus has been on understanding the effect of the active swimmers on the behavior and dynamics of passive particles. Only a few studies have considered the coupled dynamics of an active swimmer interacting with a passive particle, revealing various pairing-induced motion - circular, straight and slalom, in both experiments and simulations\cite{ishikawa2022pairing,ishikawa2025simple}. Therefore here, we study the less explored side of the interaction problem i.e. how does the presence of passive particles influence the dynamics of the active swimmer itself? \\
 
In this work, we performed experiments of Marangoni swimmers interacting with interfacial passive particles inside an annular channel geometry. The dynamics of the swimmer with increasing passive particle concentration was investigated by tracking the swimmer from recorded videos. The variation of the swimmer's speed  was studied as a function of the particle concentration. Finally, a simple analytical model, describing the swimmer-raft system, was proposed to explain the trend in the swimmer speed.

\section{Material and Methods} \label{Materials and Methods}

\begin{figure}[ht]
\centering
\includegraphics[width=\textwidth] 
 {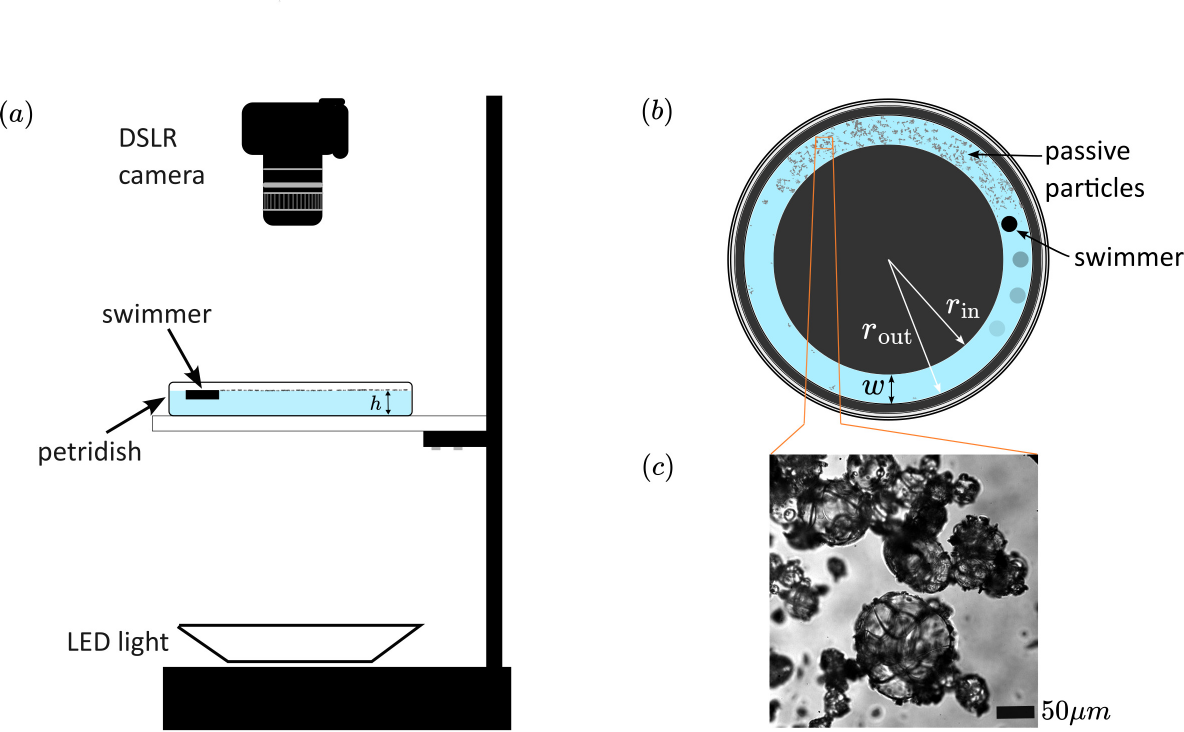}
\caption{ (a) A schematic of the experimental setup used for recording swimmer motion. The petridish was positioned on an acrylic sheet to permit illumination from below. The tiny gray spheres at the air-water interface are the passive particles;(b)Top view schematic of the petridish illustrating swimmer interacting with the passive particles in the annular channel geometry. The swimmer is moving in the counterclockwise direction. Circles in varying shades of gray indicate the swimmer’s previous positions in the channel. $w$ is the width of the annular channel. $r_{\textrm{out}}$ is the radius of the outer ring and $r_{\textrm{in}}$ is the radius of the inner circular wall forming the annular channel.; (c) Optical microscope image showing the passive particles (hollow glass microspheres). }
\label{figure_expt_setup}
\end{figure}

\subsection{Swimmer fabrication} 
\label{swimmer_prep}
Camphor-agarose disks were used as active swimmers in the experiments. Based on previous studies \cite{boniface2019self,soh2008dynamic}, these disk-shaped swimmers were prepared one day prior to the experiments following a multi-step protocol. First, an agarose (Benchmark Scientific, Inc.) in DI water solution (5$\%$ w/w) was prepared and heated to boiling. It was subsequently poured into a SLA-printed (Form 3+, FormLabs) rectangular pool of depth 0.5 mm (Fig. S3). Excess liquid was skimmed off using a scraper for a uniform layer height. Once cooled, the liquid formed an agar gel sheet of thickness $\sim$ 0.5 mm. Disk-shaped swimmers of diameter 4 mm were punched out of the sheet using a biopsy puncher (Miltex Biopsy Punch with plunger, TED PELLA, Inc.). Next, the punched out swimmers were submerged in a pure methanol solution (99.8$\%$, Sigma-Aldrich, Co.) for at least 2 hours. Following that, the swimmers were transferred to a methanol solution saturated with camphor (96$\%$, Sigma-Aldrich, Co.) i.e. 1.1 g/ml and kept for at least 12 hrs. Finally, approximately 5 mins before experiments, the swimmers were soaked in water saturated with camphor (1.2 g/l) for at least 1 min to induce the camphor to precipitate inside the agar gel matrix. This turned the swimmer completely white and opaque, making it ready for experiments. If the experiments were not performed after 12 hrs of submerging the swimmers in the methanol solution saturated with camphor, they were stored in the same solution for a maximum of 24 hours.

\subsection{Experimental protocol}

The swimmer experiments were performed within an annular channel (see \ref{Appendix annular channel} for reasoning behind the choice of geometry) formed inside a circular polystyrene Petri dish (VWR International, LLC; diameter = 140 mm) (Figs. \ref{figure_expt_setup}a,b). The annular channel was created using two floating polycarbonate films (McMaster-Carr, thickness = 0.005 inch): a circular ring as the outer wall of the channel and an inner disk as the inner wall of the channel (Fig. \ref{figure_expt_setup}b). The Petri dish was filled with deionized (DI) water. Since  speed of the swimmer is dependent on the water depth $h$, for all our experiments $h$ was fixed to approx. 20 mm, which is in the infinite-depth regime as per \cite{boniface2019self}. The swimmer motion was recorded using an overhead DSLR camera (NIKON) at 30 fps. An LED light with a diffuser was used to uniformly illuminate the Petri dish from below. For studying the interaction with passive floating particles, hollow glass microspheres (Fasco Epoxies, Inc; radius, $r_{\textrm{p}}$ $\approx$ $74 \pm 26$ $\mu$m, details in SI, Fig. S2) were gently introduced into the annular channel before the swimmer. All experiments were performed in a dark environment to improve image contrast and avoid any reflections from the background. Moreover, all experiments were performed in the same room to minimize variability due to changes in humidity and temperature.\\

Multiple experiments were performed sequentially for each swimmer to measure different aspects of its dynamics. Prior to each trial, the amount of passive particles was precisely weighed using a benchtop weighing scale (Mettler Toledo). First, before the swimmer was introduced at the air-water interface, a recording of the channel uniformly filled with hollow glass microspheres was captured using a macro lens (NIKKOR 105 mm, 1:2.8D) to measure the initial particle concentration ($\phi_{\textrm{ini}}$), henceforth referred to as the initial packing fraction. Here, packing fraction was defined as the fraction of the interface area covered by the passive particles. Next, the swimmer was quickly cleaned ($\sim$ 5 secs) in a Petri dish filled with DI water to remove any camphor particles adhered to its surface. Then, the swimmer was introduced into a Petri dish with no passive particles at the interface and the swimmer motion was recorded for 30 secs. Following this step, the swimmer was transferred to the Petri dish with passive particles,  where its motion was recorded for 1 minute.  A second video of the swimmer motion in the same Petri dish was recorded with a macro lens (NIKKOR 105 mm, 1:2.8D) to measure the mean packing fraction in the raft ($\phi_{\textrm{raft}}$) infront of the swimmer. Finally, the swimmer was transferred back to the Petri dish with no passive particles and its motion was recorded again for 15 secs. Since, the swimmer speed is known to gradually decrease over time (\cite{boniface2019self}), the total experimental duration for each swimmer was kept under 3 mins. \\

While transferring the swimmer between different Petri dishes, a clean metal spatula was used to avoid any contamination. After all experiments with each swimmer, all Petri dishes were thoroughly cleaned and then refilled with DI water in order to minimize any effects of contamination of the interface due to remnant camphor particles or dust. Cases where the swimmer would not self-propel without any particle and even after the Petri dish had been cleaned and refilled  were not considered for analysis. Following the above protocols, a total of 44 swimmers were tested for this study.

\subsection{Image analysis}
The swimmer tracking, from the recorded videos, was performed using a custom MATLAB code based on masking and circle detection algorithms. The particle packing fraction was measured using a custom MATLAB code that first binarized the images based on a fixed intensity threshold and then counted the black pixels, representing particles, within the annular region of interest (Fig. \ref{fig_vel_phi_avg}a, Details in \ref{Appendix packing fraction}). The individual particles were not resolved in the images, so within each detected black pixel the particles were assumed to be at random close packing i.e. $\phi=0.82$ \cite{berryman1983random}. Hence, a correction factor of 0.82  was applied to the packing fraction directly obtained from the image analysis. The passive particles within the raft were tracked using PIVLab app\cite{thielicke2021particle} on MATLAB. For experiments with very low packing fraction ($\phi_{\textrm{ini}}<0.1$), few particles within the raft were tracked using the DLTdv8 app on MATLAB \cite{hedrick2008software} to measure the mean speed (\ref{Appendix DLTdv}).


\section{Results and discussion} \label{Results and discussion}
\subsection{Swimmer dynamics}

\begin{figure}[htbp]
    \centering
\includegraphics[width=\textwidth,trim={0 1.5cm 0 0},clip]{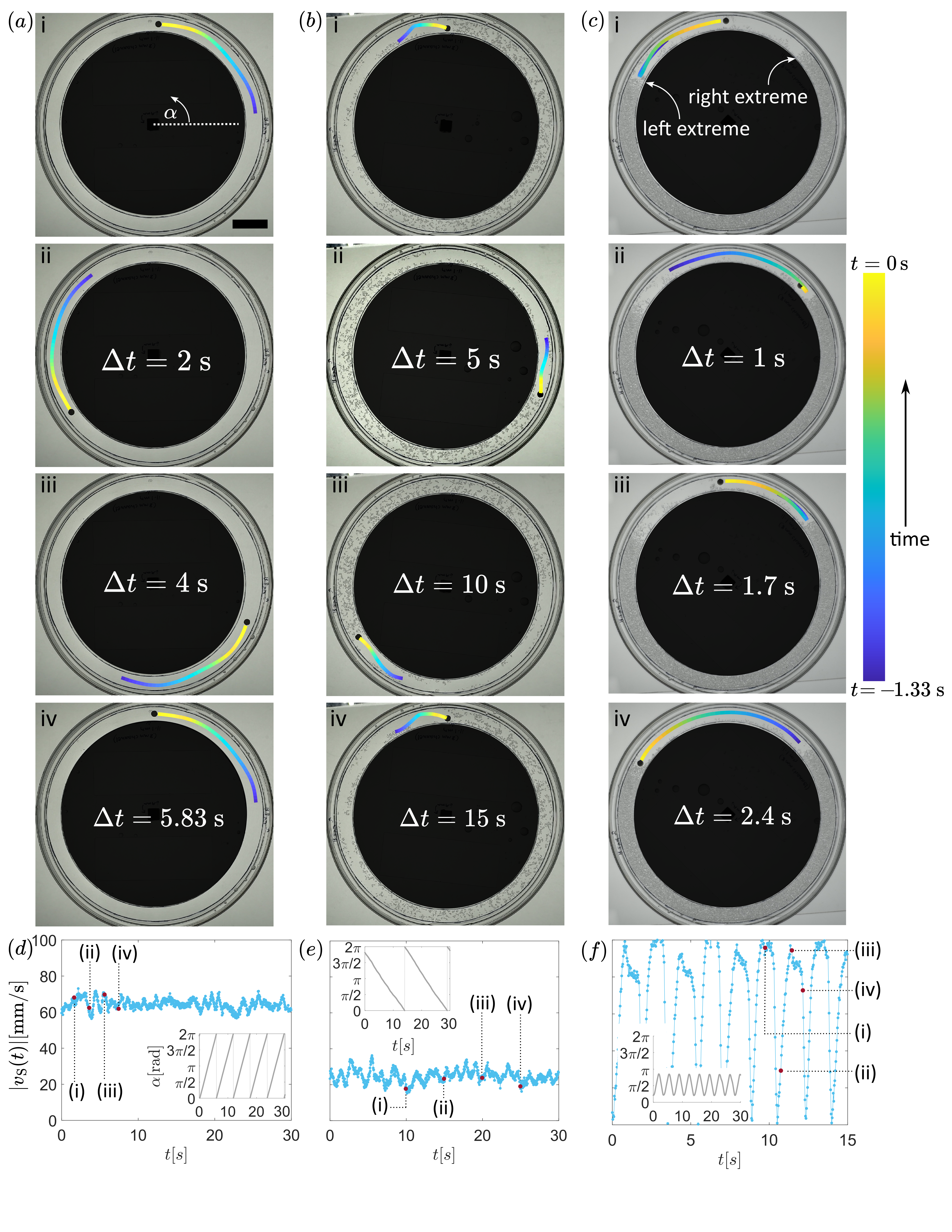}
    \caption{ (a-c) Representative sequential images of the swimmer motion in the annular channel with (a) $\phi_{\textrm{ini}}=0$ (steady uni-directional motion at mean speed $v_{\textrm{s}}\approx 65$ mm/s), (b)  $\phi_{\textrm{ini}} \approx 0.35$ (steady uni-directional motion at mean speed $v_{\textrm{s}}\approx 25$ mm/s) and (c) $\phi_{\textrm{ini}} \approx 0.6$ (oscillatory motion). $\Delta t$ is the time difference from the first frame (i) in each image sequence. The colored line in the channel represents the swimmer trajectory over the last 1.33s; (d-f) Time evolution of the swimmer speed corresponding to the cases shown in (a–c), respectively. Insets show the time evolution of the swimmer's azimuthal position ($\alpha$). The scalebar is 16 mm.}
    
    \label{figure_swimmer_dynamics}
    
\end{figure}

 When the camphor disk (referred to as the swimmer later on) was introduced into a channel uniformly filled with certain initial passive particles ($\phi_{\textrm{ini}}$), the particles first got more packed due to the sudden radially outward Marangoni flow setup at the interface by the swimmer.  Within a few secs, the swimmer started steady (with small fluctuations) uni-directional motion at a mean azimuthal speed $v_{\textrm{s}}(\phi)$ (Fig. \ref{figure_swimmer_dynamics}b,e, movie S2), hereafter referred as speed of the swimmer, which was slower than the mean speed of the same swimmer in the channel without any particles $v_{\textrm{s}}(\phi=0)$. \\

Up to a critical packing fraction $\phi_{\textrm{ini}} \approx 0.45$, a steady uni-directional motion was observed with the mean swimmer speed $v_{\textrm{s}}(\phi)$ decreasing with increasing packing fraction ($\phi_{\textrm{ini}}$). Note that the speed of the swimmer below critical packing fraction was never perfectly steady and has small fluctuations due to wall hitting and radial motion (Fig. \ref{figure_swimmer_dynamics}e), yet we still refer to it as steady to differentiate from the oscillatory behavior observed above the critical packing fraction.\\

For $\phi_\textrm{ini} \gtrsim 0.45$, the swimmer motion transitioned from steady uni-directional motion to oscillatory motion in the channel ( Fig.~\ref{figure_swimmer_dynamics}c, Inset in Fig. \ref{figure_swimmer_dynamics}f, movie S3). More precisely, the swimmer was trapped between the two ends of the raft of particles and performed back and forth motion between these extreme positions (Fig. \ref{figure_swimmer_dynamics}c) in the channel. During the oscillatory behavior, the swimmer's peak speed was close to the mean speed of the swimmer in the absence of particles. This is attributed to the annular channel being almost free of any particles in-between the two extreme points where the swimmer undergoes back and forth motion. Although not explored in this study in detail, for packing fractions approaching the random close packing in 2D i.e. $\phi_{\textrm{ini}} \approx 0.82$, the swimmer motion became dominantly stationary with sporadic intervals of motion (Fig. S4, movie S4).

\subsection{Swimmer-raft interaction in steady regime}
Now, we will focus on understanding the motion of the swimmer in the steady regime. Since the swimmer self-propels by generating local surface tension gradients, one might expect its interaction with the passive particles to be limited to short range. However, the 1D confinement in the annular channel resulted in the interaction to be quite long range. Consequently, while the swimmer performed steady uni-directional motion, it not only displaced the particles nearby, but transported majority of the particles in the channel in the same direction, effectively pushing them like cargo (movie S5). \\

\begin{figure}[htbp]
\centering
\includegraphics[width=\textwidth,trim={0 6cm 0 0},clip] 
 {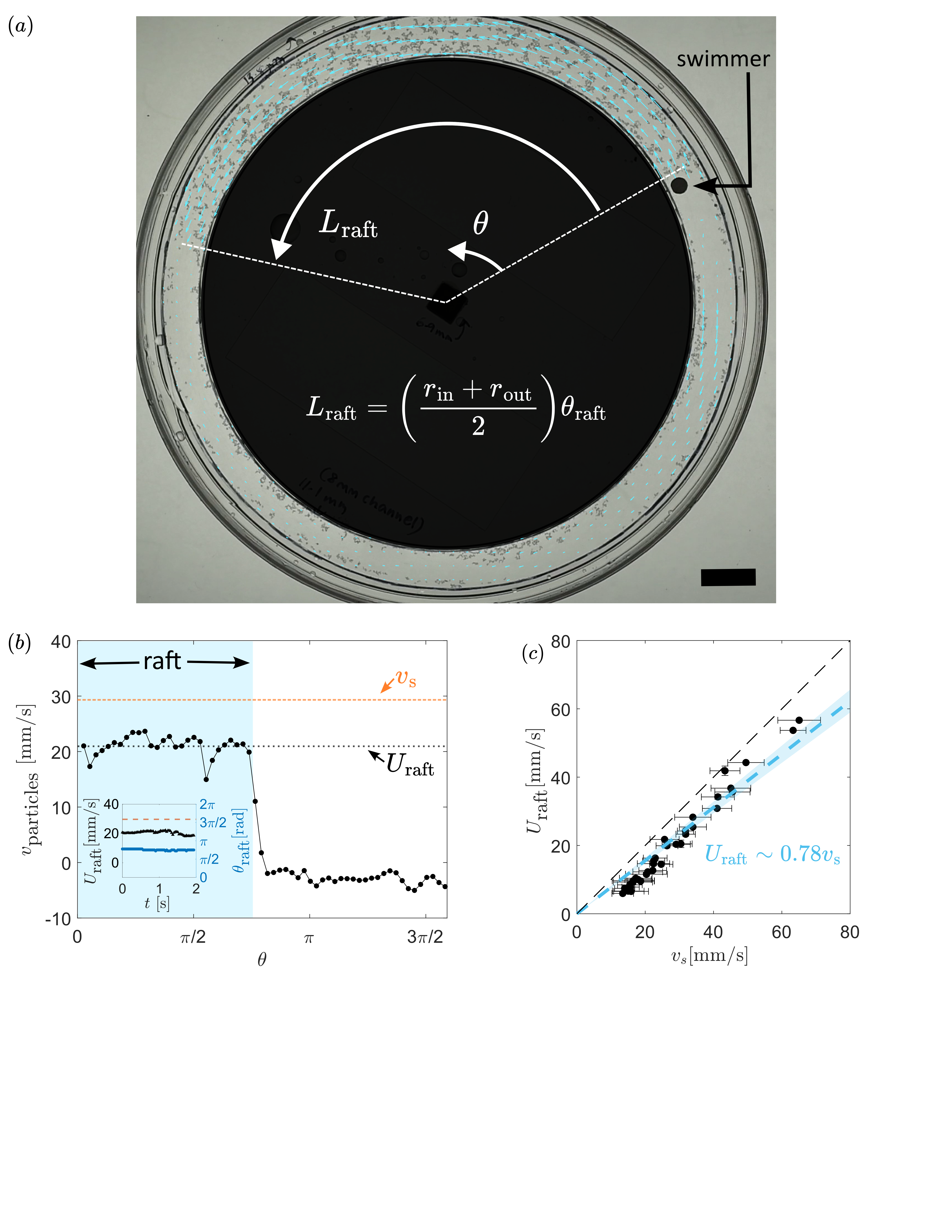}
\caption{ (a) Image illustrating the particle raft being pushed by the swimmer. Swimmer is moving in the counter-clockwise direction. The blue arrows are local velocity vectors of the particles obtained from PIV. Scalebar is 8 mm;(b) The local speed of the particles ($v_{\textrm{particles}}$) as a function of azimuthal angle ($\theta$) measured from the swimmer, corresponding to the case shown in (a). The blue shaded portion highlights the particle raft being dragged at a speed close to the swimmer's speed ($v_{\textrm{s}}$). The dotted horizontal line in orange represents the swimmer speed. The dotted horizontal line in black is the mean speed of the particles within the raft ($U_{\textrm{raft}}$). Inset shows the time evolution of $U_{\textrm{raft}}$ and $\theta_{\textrm{raft}}$; (c) Mean speed of the raft ($U_{\textrm{raft}}$) plotted against the swimmer speed for all experiments. The black dotted line is $x=y$. The blue dashed line is a linear fit to the data. The shaded area around it represents the 95$\%$ confidence intervals.}
\label{figure_raft_properties}
\end{figure}

To quantify this observation, we performed particle image velocimetry (PIV) analysis on the passive particles in the channel to extract local particle velocities from the recorded videos. Fig. \ref{figure_raft_properties}a shows the velocity vectors (blue) from the PIV analysis for a representative case. The mean azimuthal component of the particle velocities are plotted in Fig. \ref{figure_raft_properties}b as a function of the azimuthal angle from the swimmer ($\theta$), where this mean was calculated over all velocity vectors along the width of the channel at each discrete azimuthal angle. Note that particles close to the channel walls (inner and outer) had slower velocity due to slower fluid velocity near the walls, but we ignore this for simplicity. The plot shows that the mean speed of the particles in the channel is nearly constant upto a certain azimuthal distance from the swimmer ($\theta \approx \frac{3 \pi}{4}$ for the representative case in Fig. \ref{figure_raft_properties}b). Beyond that azimuthal distance, the speed of the particles suddenly drops indicating that these are not pushed by the swimmer. Based on this observation, we henceforth refer to all particles moving with the swimmer (at nearly similar speed) as the raft. The mean speed of all the particles in the raft was defined as $U_{\textrm{raft}}$ (referred to as mean speed of raft later on). Further, $\theta_{\textrm{raft}}$ and $L_{\textrm{raft}}$ was defined as the angular length and arc length of the raft, respectively. Both $\theta_{\textrm{raft}}$ and $U_{\textrm{raft}}$ remained nearly constant over time for an individual experiment (Inset in Fig. \ref{figure_raft_properties}b). Note that particles right behind the swimmer are pushed in a direction opposite to the swimmer motion because of the outward Marangoni flow near the swimmer, Fig. \ref{figure_raft_properties}a. Furthermore, for some experiments with very low packing fraction, particle tracking was performed since PIV yielded poor results due to void regions in the channel (\ref{Appendix DLTdv}). \\

Finally, the mean speed  of the raft is plotted against the mean swimmer speed for all experiments in Fig. \ref{figure_raft_properties}c. It indicates that although the mean speed of the raft is linearly proportional to the swimmer speed, it has a smaller magnitude (i.e.  $U_{\textrm{raft}} \sim 0.78 v_{\textrm{s}}$). This is due to the local leakiness of the raft described earlier.\\

\begin{figure}[ht]
\centering
\includegraphics[width=\textwidth,trim={0 21cm 0 0},clip] 
 {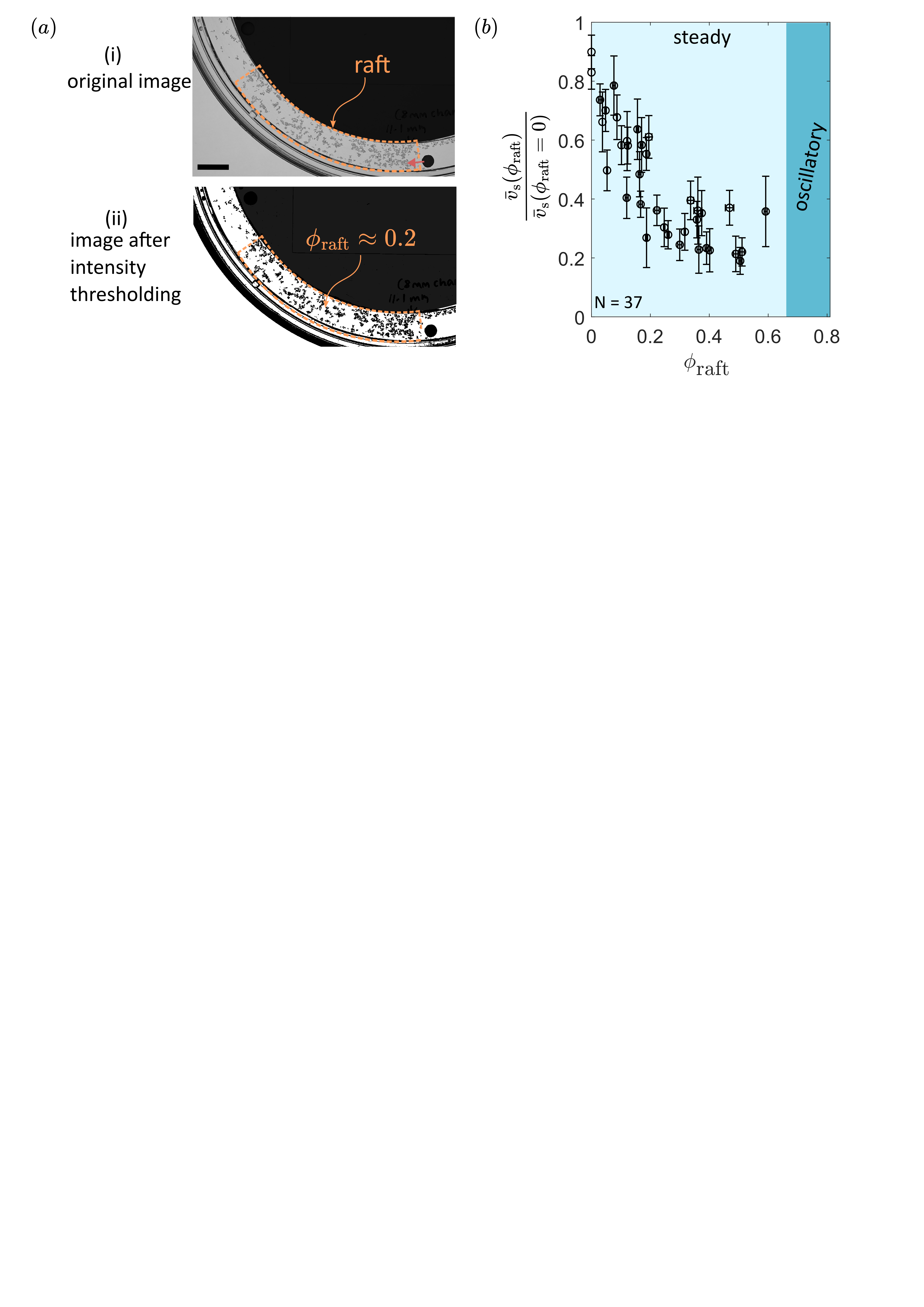}
\caption{ (a) Binarized image (below) obtained from the original image (above) for calculating the mean packing fraction in the particle raft ($\phi_{\textrm{raft}}$). Scalebar is 8 mm. (b) Variation of the rescaled swimmer speed with the mean packing fraction of the particle raft ($\phi_{\textrm{raft}}$). Each data point corresponds to an individual experiment. $N=37$ indicates the total number of data points in the steady regime. The light blue shaded portion indicates steady regime i.e. unidirectional motion of the swimmer, while the darker blue indicates oscillatory regime. The errorbars are standard deviation values.}
\label{fig_vel_phi_avg}
\end{figure}

We then explored the speed dependence of the swimmer on the packing fraction in the steady regime. The mean swimmer speed is not uniform across swimmers even without any particles at the interface, potentially due to irregularities in the geometry or variations in the thickness during the agar disk fabrication. Therefore, to remove the effects pertaining to the fabrication process, here the mean speed of each swimmer in the presence of particles ($v_{\textrm{s}}$) was rescaled with its mean speed without any particles ($v_{\textrm{s}}(\phi=0)$). In Fig. \ref{fig_vel_phi_avg}b, the rescaled speed of the swimmer is plotted against mean packing fraction of the raft ($\phi_{\textrm{raft}}$), illustrated in Fig. \ref{fig_vel_phi_avg}a. It reveals that the rescaled speed monotonically decreases with increasing $\phi_{\textrm{raft}}$. The decrease is faster for lower packing fractions than higher values close to the transition to oscillatory mode, indicating a non-linear dependence. Here, the deviation for mean packing fraction of the raft is measured as its standard deviation over time ($\sim 2-4$ secs). Note that here the mean packing fraction of the raft ($\phi_{\textrm{raft}}$) is chosen as the horizontal axis over initial packing fraction ($\phi_{\textrm{ini}}$) since it provides more direct knowledge of the swimmer-raft interaction during self-propulsion.

\subsection{Analytical model for swimmer speed}
 
In order to explain the velocity dependence of the swimmer on the packing fraction in the steady regime (Fig. \ref{fig_vel_phi_avg}), we present a simple model for the interacting swimmer-raft system.
\subsubsection{No particles at interface}
\begin{figure}[ht]
\centering
\includegraphics[width=\textwidth] 
 {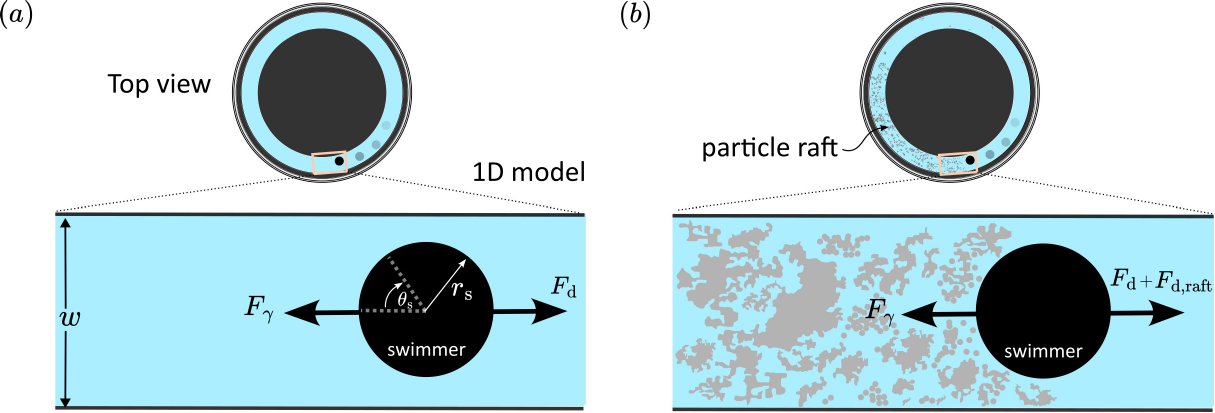}
\caption{ A schematic for the idealized 1D model of the swimmer within confined channels (a) without particles and (b) with particles.}
\label{fig_schematic_model}
\end{figure}

 First, a set of equations which describe the motion of the swimmer in the absence of any particles, based on the point-source model for isotropic swimmers\cite{boniface2019self,gouiller2021two,suematsu2014quantitative}, are presented. For modeling purpose, an idealized 1D geometry, neglecting curvature effects in the real geometry (annular channel) is considered (Fig. \ref{fig_schematic_model}a). The motion of the Marangoni swimmer is governed by two coupled equations :  
 1) Force balance between surface tension ($F_{\gamma}$) and viscous drag on the swimmer ($F_{\textrm{d}}$), assuming steady state motion 
\begin{equation}
    F_{\gamma}=F_{\textrm{d}},
    \label{force balance eqn}
\end{equation}
where $F_{\gamma}=\oint \gamma(c) dl$ and $F_{\textrm{d}} = \frac{16}{3}\mu r_{\textrm{s}}v_{\textrm{s}}(\phi_{\textrm{ini}}=0)$ \cite{boniface2019self}, neglecting lateral confinement effects.
2) Convection-diffusion equation for the camphor at the interface
\begin{equation}
    \frac{dc}{dt}+\vec{u} \cdot \nabla c=D\nabla^2 c -k c + J \delta(\vec{r}),
    \label{convec-diff eqn}
\end{equation}
where $c$ is the concentration of camphor at the air-water interface, $\vec{u}$ is the velocity field of water at the interface, $D$ is the diffusion coefficient of camphor in water, $k$ is the rate of sublimation of camphor at the interface, $J$ is the flux of camphor release from swimmer, $\vec{r}$ is the radial vector from the centre of the camphor disk.
Since, here the Peclet number ($Pe=\frac{r_\textrm{s} v_\textrm{s}}{2 D} \sim 10^5$) is quite high, the transport of camphor is advection dominated.  In the frame of the moving swimmer, assuming constant speed $v_{\textrm{s}}$, the convection-diffusion equation simplifies to 
\begin{equation}
    -v_{\textrm{s}} \cdot \nabla c=D\nabla^2 c -k c + J \delta(\vec{r}),
    \label{convec-diff swimmer frame}
\end{equation}
Further, an order-of-magnitude analysis revealed that the sublimation term is negligible compared to the advection term due to swimmer motion in eqn. \ref{convec-diff swimmer frame}
\begin{equation}
    \frac{[k c]}{[-v_{\textrm{s}} \cdot \nabla c]} \approx 10^{-4} \,,
    \label{scaling}
\end{equation}
where $k= 1.8 \times 10^{-2}$ s$^{-1}$ \cite{suematsu2014quantitative}. Finally, the equation $-v_{\textrm{s}} \cdot \nabla c=D\nabla^2 c + J \delta(\vec{r})$ is analytically solvable based on the point-source model for a symmetric swimmer moving at a constant speed\cite{boniface2019self}. The concentration distribution around the swimmer is given by $c(r,\theta_{\textrm{s}}) = \frac{J}{4 \pi r D} \textrm{exp} \left[\frac{-v_{\textrm{s}} r }{2 D} (1+\cos \theta_{\textrm{s}})\right]$ \cite{boniface2019self}, where $\theta_{\textrm{s}}$ is the angular position in the azimuthal direction (Fig. \ref{fig_schematic_model}a). Note the concentration distribution is independent of time since we assumed the swimmer speed to be a constant for modeling purpose, although not perfectly true in experiments. \\

Then, the net surface tension force on the swimmer can be calculated using 
\begin{equation}
F_{\gamma}=\oint \gamma dl = -2 \kappa r_{\textrm{s}}\int_0^\pi c(r_{\textrm{s}},\theta_{\textrm{s}})\cos \theta_{\textrm{s}} d\theta_{\textrm{s}} \, ,
\end{equation}
\cite{boniface2019self}  where $\kappa$ is the concentration dependence of surface tension. Substituting for $c(r_{\textrm{s}},\theta_{\textrm{s}}$) and integrating, $F_{\gamma}=\frac{\kappa J}{2 D} e^{-Pe}I_1(Pe)$. Further replacing the Bessel function with its asymptotic behavior at $Pe \to \infty$, since $Pe$ is very high in experiments, the expression for surface tension force on the swimmer simplifes to 
\begin{equation}
    F_{\gamma} = \frac{\kappa J}{2 D} \sqrt{\frac{D}{\pi r_{\textrm{s}} v_{\textrm{s}}(\phi_{\textrm{ini}}=0)}} \,.
\end{equation}

\subsubsection{With passive particles at the interface}\label{model with particles}
First, the presence of passive particles at the interface (Fig. \ref{fig_schematic_model}b) necessitates a reformulation of the swimmer's force balance. The force balance equation on a swimmer pushing the particle raft in the steady regime is Eq. \ref{force balance eqn} modified with a reaction force term from the interaction with the raft i.e.
\begin{equation}
    F_{\gamma}=F_{\textrm{d}}+F_{\textrm{d,raft}},
    \label{force_balance_with_particles}
\end{equation}
where $F_{\textrm{d,raft}}$ is the reaction force on the swimmer from the raft which is equivalent to the viscous drag experienced by the raft. Here, $F_{\gamma}$ and $F_{\textrm{d}}$ have the same expression as described in previous section, with only the velocity $v_{\textrm{s}}(\phi_{\textrm{ini}}=0)$ replaced with $v_{\textrm{s}}(\phi_{\textrm{ini}})$, which is the speed of the swimmer in presence of particles. Any dependence of the sublimation coefficient, $k$ on packing fraction is ignored in the surface tension force term, since, although the particles might reduce the rate of sublimation from the interface, the contribution of sublimation term to the camphor concentration distribution is negligible compared to advection, as revealed in eqn.\ref{scaling}. Moreover, the brownian diffusion of camphor molecules is not expected to be affected due to presence of passive particles (micrometer scale) which are orders of magnitude larger than the diffusing molecules (nanometer scale) itself. Also, we assume that the flux of camphor release ($J$) is constant for a swimmer (given that the total experiment time for each swimmer is much smaller than the timescale of reduction in $J$ \cite{boniface2019self}) and unaffected by the presence of particles at the interface.\\

\begin{figure}[ht]
\centering
\includegraphics[width=\textwidth] 
 {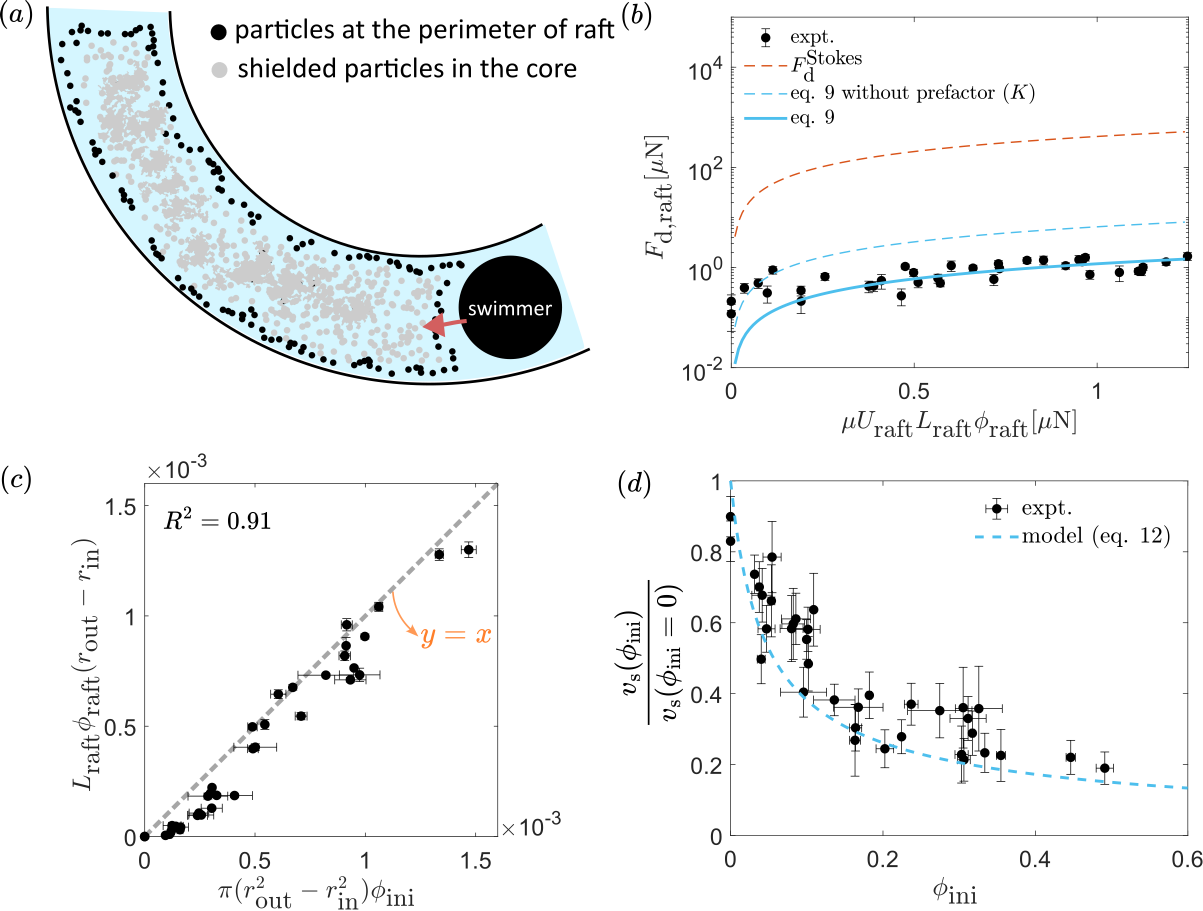}
\caption{ (a) A schematic illustrating the shielding effect on the particles in the raft; (b) Comparison of different drag models for the particle raft. The black dots represent the experimental expectation of the drag force. The errorbars represent standard deviation. The blue line is the drag estimated using $6K\pi \mu \alpha_{||} U_{\textrm{raft}}\phi_{\textrm{raft}}L_{\textrm{raft}}$, while the blue dotted line is without including the prefactor ($K$). The orange dotted line represents Stokes drag estimated without incorporating shielding effect; (c) Experimental validation of the number conservation of passive particles in the raft i.e. eqn. \ref{mass_conservation}; (d) Comparison between model predictions i.e. eqn. \ref{final_vel_eqn} and experimental data representing the variation of rescaled velocity of the swimmer with initial packing fraction. The errorbars represent standard deviation.}
\label{fig_model_results}
\end{figure}

Next, a model for the viscous drag on the raft is proposed. The Reynolds number ($Re_{\textrm{p}}={2\rho U_{\textrm{raft}} r_{\textrm{p}}}/{\mu}$) of individual passive particle was approximately 1, where $\rho$ is density of water. Hence, drag on each particle in the raft was estimated using Stokes drag for a sphere moving at air-water interface i.e. $ 6 \pi \alpha_\parallel \mu r_{\textrm{p}} U_{\textrm{raft}}$  \cite{villa2020motion,dorr2016drag}. Here,  $\alpha_\parallel=({8}/{9 \pi})\sin \theta_{\textrm{p}} \left[1+(4/3\pi) \cot(\theta_{\textrm{p}}/2) \right]$ is the pre-factor dependent on the contact angle of the particle ($\theta_{\textrm{p}}$, details in SI, Fig. S1). Previous studies on rafts or aggregates of particles (with $Re_{\textrm{p}}\sim1$) estimated the total drag by either summing the drag on individual particles \cite{vassileva2006restructuring} or by calculating the drag on the effective sphere formed by the aggregate \cite{lagarde2019capillary,lagarde2020probing}. However, in all these scenarios, the total particle count in the rafts were less than 400, resulting in raft sizes on the order of millimetres. In contrast, the rafts in this study are at least an order of magnitude larger in size. Consequently, a different modeling approach is adopted here. It is worth noting that the particle raft in our experiments can be viewed as a cluster of many spheres with void spaces in between, described by the packing fraction. Previous studies on the drag on a cluster of particles in Stokes flow have shown that the drag on the particles at the perimeter are much higher than the particles in the core of the cluster \cite{filippov2000drag}. In other words, the particles in the core experience a shielding effect due to a reduced local flow speed. As a first order approximation of this shielding effect, we here assume that the drag on the raft is dominated by the drag contribution on the particles at the perimeter only, as illustrated in schematic \ref{fig_model_results}(a). Under this approximation, the total drag on the raft becomes drag on each particle times the number of particles at the perimeter of the raft ($N_{\textrm{peri}}$). The number of particles at the perimeter of the raft was estimated as 
\begin{equation}
N_{\textrm{peri}}= \phi_{\textrm{raft}}\frac{(L_{\textrm{raft}}+w)}{r_{\textrm{p}}}.
\end{equation}
This leads to the total drag force on the raft given by
$F_{\textrm{d,raft}}= 6 \pi \mu \alpha_\parallel U_{\textrm{raft}} \phi_{\textrm{raft}} (L_{\textrm{raft}}+w)$.\\

Further assuming that $w \ll L_{\textrm{raft}}$, based on experiments, total drag on raft simplies to $F_{\textrm{d,raft}}= 6 \pi \mu \alpha_\parallel U_{\textrm{raft}} \phi_{\textrm{raft}} L_{\textrm{raft}}$. This is plotted in Fig. \ref{fig_model_results}b as the blue dotted line. It indicates that the model for the drag on the raft follows similar trend but overestimates the experimental expectation. The experimental estimation of drag on the raft is based on the swimmer motion from videos and force balance (\ref{Appendix F_d_raft expt}). Potential reasons for the overestimation are - (a) drag on the particles at the perimeter is less than the theoretical estimate from Stokes drag due to walls on one side, (b) passive particles are not perfect spheres, thus  experience less drag, (c) Inaccuracy in the image based measurement of the contact angle due to obscured downward curvature of the meniscus near the passive particle and (d) deviation of theoretical estimate of $\alpha_\parallel$ from reality. Hence, a prefactor ($K$) is incorporated resulting in 
\begin{equation}
F_{\textrm{d,raft}}= 6 K \pi \mu \alpha_\parallel U_{\textrm{raft}} \phi_{\textrm{raft}} L_{\textrm{raft}}.
\end{equation}
The prefactor is calculated by fitting the theoretical drag law to the experimental estimation (Fig. S5). The best fit gives $K$ = $0.18 \pm 0.01$. The solid blue line in Fig. \ref{fig_model_results}b is the theoretical expression for drag on raft with the prefactor included. Also, note that the drag on the raft based on Stokes drag on all particles within the raft ($F_{d}^{\textrm{Stokes}}= 6\mu \alpha_{||}U_{\textrm{raft}}L_{\textrm{raft}}\phi_{\textrm{raft}}w / r_{\textrm{p}}$) overestimates the experimental estimation by two orders of magnitude, plotted as orange dotted line in Fig. \ref{fig_model_results}b. This is a justification for the shielding effect assumption used in the model above. \\

Second, an equation relating the speed of the raft with the speed of the swimmer is given by 
\begin{equation}
    U_{\textrm{raft}}=C v_{\textrm{s}}.
    \label{raft_swimmer_speed_relation}
\end{equation}
since the swimmer pushes the raft in the annular channel. The coefficient $C$ was emperically set as $0.78 \pm 0.04$  (95$\%$ confidence levels) based on experimental measurements, shown in Fig. \ref{figure_raft_properties}c. Lastly, a conservation equation for the total number of passive particles in the raft, assuming all passive particles introduced to the interface at the beginning of the swimming experiment stays within the raft, is given by 
\begin{equation}
    \pi (r^2_{\textrm{out}} -r^2_{\textrm{in}}) \phi_{\textrm{ini}} = L_{\textrm{raft}} \phi_{\textrm{raft}} (r_{\textrm{out}} - r_{\textrm{in}}) \,.
    \label{mass_conservation}
\end{equation}
The rationale behind this assumption is the observation of small amount of particles in the channel outside the raft compared to the total number of particles inside the raft (movie S2).
But, since experiments indicated some leakage in the proximity of the swimmer (movie S2, S6), the number conservation equation is not perfect. Nevertheless, Fig. \ref{fig_model_results}c indicates that based on experimental measurements it is a reasonable assumption. \\

Finally, a theoretical expression for the speed of the swimmer in the presence of passive particles as a function of the two initial conditions for the system, i.e. $v_{\textrm{s}}(\phi_{\textrm{ini}}=0)$ and $\phi_{\textrm{ini}}$, was calculated by solving eqns. \ref{force balance eqn}, \ref{force_balance_with_particles}, \ref{raft_swimmer_speed_relation}, and \ref{mass_conservation}, namely 
\begin{equation}
    v_{\textrm{s}}(\phi_{\textrm{ini}})=v_{\textrm{s}}(\phi_{\textrm{ini}}=0) \left [\frac{\frac{16}{3}r_{\textrm{s}}}{\frac{16}{3}r_{\textrm{s}} + 4.2 \pi^2 \alpha_\parallel K (r_{\textrm{in}}+r_{\textrm{out}}) \phi_{\textrm{ini}} } \right]^{2/3}.
    \label{final_vel_eqn}
\end{equation}
A comparison of the expression from the model and experiments is presented in Fig. \ref{fig_model_results}d. The simple first order model captures the overall trend in the speed of the swimmer from experiments. The deviation of the experimental data from the theoretical estimate roots from secondary effects such as radial motion of swimmer, intermittent wall hitting and local leakiness of particle raft, which are not incorporated in the model presented in this study for simplicity.

\section{Conclusion} \label{Conclusion}
In order to understand the effects of interaction of an active swimmer with passive particles in its environment on its self-propulsion, in this study we performed experiments of camphor-agarose swimmers interacting with hollow glass microspheres at the air-water interface within the confinement of annular channel. The experiments revealed two distinct behaviors of the swimmer (Fig. \ref{figure_swimmer_dynamics}) in presence of different initial particle packing fraction ($\phi_{\textrm{ini}}$). For $\phi_{\textrm{ini}} \lesssim 0.45$, the swimmer motion was nearly a steady uni-directional motion. The steady speed of the swimmer decreased with increasing $\phi_{\textrm{ini}}$. On the other hand, for  $\phi_{\textrm{ini}} \gtrsim 0.45$ the swimmer motion transitioned to oscillatory type, where the swimmer performed to and fro motion between the two ends of the raft. The major observation from experiments in the steady regime was that under the confinement of the annular channel the particle raft is pushed by the swimmer along with it, slowing down the swimmer.  While, in the oscillatory case, the
swimmer goes back and forth in a particle-free zone except near the two ends of the region. Consequently, the peak speed during the oscillatory behavior was close to the mean speed of the swimmers without any particles at the interface. Moreover, the swimmer increased the packing of the particles as it approached either ends of the raft, simultaneously  slowing down and eventually turning around to propel in the opposite direction. While, as the swimmer propelled away the particles in the raft gradually relaxed from a packed state.
 \\

In the steady regime, the particle raft speed was linearly proportional to the swimmer speed but smaller by a factor of 0.78. This reduction was attributed to the local leakiness of the raft near the swimmer. A plot of the rescaled speed of the swimmers against the mean packing fraction of the raft revealed a monotonic decreasing trend (Fig. \ref{fig_vel_phi_avg}b). Further, it highlighted a faster decrease at low packing fraction and a slower decrease at higher packing fractions close to the transition to oscillatory mode. \\

 Finally, we proposed a simple model to explain the swimmer's speed dependence on packing fraction in the steady regime. The model solves the force balance on the swimmer under a 1D assumption, while incorporating velocity constraints between the swimmer and raft as well as conserving the number of particles at the interface. The swimmer-raft interaction was captured as a reaction force term, equivalent to the viscous drag on the raft,  in the force balance equation on the swimmer. Due to shielding effect on the particles in the raft, particles at the perimeter experience significantly higher drag compared to the ones in the core. Hence, as a first order approximation, here we considered drag from only perimeter particles. Although, the theoretical drag follows a similar trend as the prediction from experiments, it overestimates the value. A prefactor was used to capture this. Overall, the final theoretical estimate of the speed of the swimmer (Eqn. \ref{final_vel_eqn}) matches the general trend of the experimental data (Fig. \ref{fig_model_results}d).\\
 
 Although the model presented here captures the general trend of the swimmer speed, there is some deviation of experiments from the model. This is potentially because it ignores 2D effects such as radial directional motion and collisions with the channel boundary. Also, a rigorous model of the shielding effect on the drag, considering the variation in the extent of shielding based on the particle location within the raft, could likely explain the prefactor. This calls for future studies to build on the model presented here. Moreover, the regime transition to oscillatory behavior needs to be investigated to understand the critical condition. Further, from an experimental perspective, it would be interesting to study the dynamics and interactions with passive particles in a two-dimensional pool. A recent work highlights the richness of the dynamics even without passive particles \cite{cruz2024active}. Also, the diverse emergent behaviors when multiple swimmers interact with particles is worth exploring. Overall, this article highlights the cargo transport potential of Marangoni swimmers under 1D confinement and inspires further work on designing active swimmers for controlled manipulation of cargo.

\section{Appendix} \label{Appendix}

\begin{figure}[ht]
\centering
\includegraphics[width=\textwidth] 
 {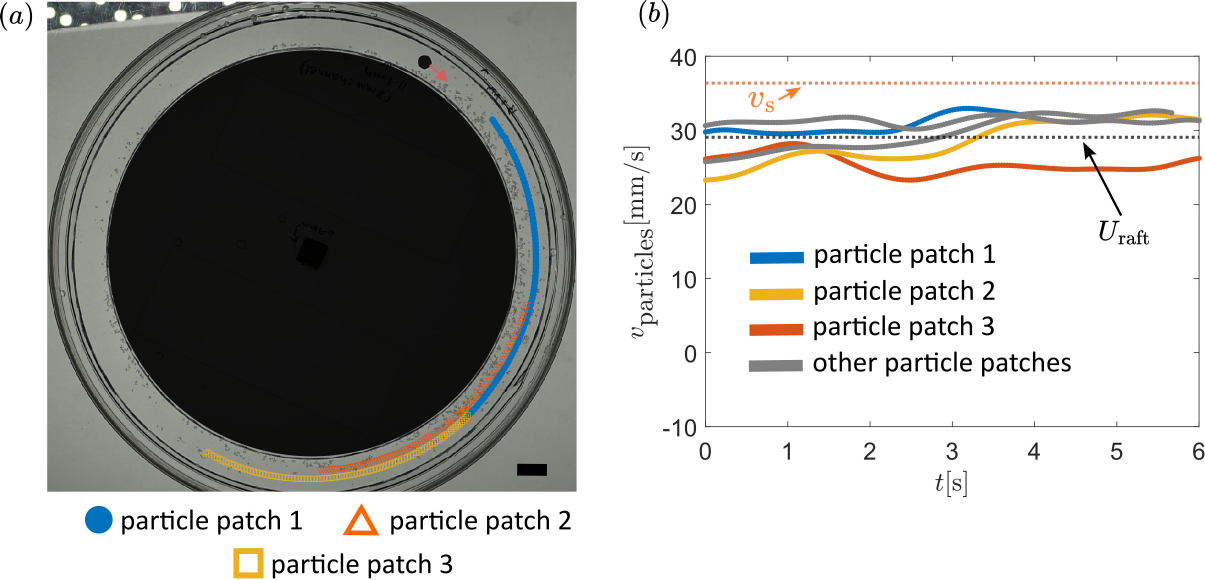}
\caption{ (a) Trajectory of three representative particle patches for 3 secs obtained from dltdv8a. Scalebar is 8 mm ;(b)Temporal evolution of the speed of multiple particle patches tracked for the experiment shown in (a). The orange dotted line represents the swimmer speed. The black dotted line represents the mean speed of the raft, computed as average speed of the tracked patches. }
\label{figure_dltdv_track}
\end{figure}

\subsection{Reasoning behind choice of annular channel geometry}\label{Appendix annular channel}
In a circular geometry, the motion of the swimmer is known to be random with some dominant behaviors, such as circular motion along the boundary and active chiral motion when away from the boundary \cite{cruz2024active}. Instead, when confined in an one-dimensional infinite channel (annular channel) the swimmer moves at a steady speed in a particular direction (Fig. \ref{figure_swimmer_dynamics}a,d, Inset in Fig. \ref{figure_swimmer_dynamics}d, movie S1), which is randomly picked at the moment the swimmer was introduced to the channel. Moreover, in the annular channel, radial direction of the swimmer was negligible (Fig. \ref{figure_swimmer_dynamics}a, movie S1). Therefore, due to a more controlled behavior and a single dominant degree of freedom in the annular channel it was chosen over a circular channel, for understanding the interaction with passive particles at the air-water interface.

\subsection{Particle tracking using DLTdv8a for low packing fraction} \label{Appendix DLTdv}
In order to calculate the mean speed of the raft ($U_{\textrm{raft}}$) for very low packing fraction experiments, individual particle patches within the raft were tracked from the videos using a MATLAB based app, DLTdv8a \cite{hedrick2008software} (Fig. \ref{figure_dltdv_track}a). For each video, multiple particle patches, located at different azimuthal positions from the swimmer, were tracked (Fig. \ref{figure_dltdv_track}b). The mean of the speeds of all patches was defined as $U_{\textrm{raft}}$ for these experiments. For this definition, we neglect the temporal variation of the speed of individual patches. Also, the deviation of speed across patches located at different azimuthal locations from the swimmer is considered small. 

\subsection{Steps of custom image analysis for packing fraction measurement} \label{Appendix packing fraction}
For calculating the packing fraction of particles in-front of the moving swimmer from the original image (\ref{fig_vel_phi_avg}a), first the two circular arcs, corresponding to the inner and outer channel walls, were detected by locating three distinct points on the arc circumference. Each point was detected by searching for intensity spikes along vertical direction at a fixed horizontal position in the image (after removing small unnecessary objects in the original image). Once detected, the portion of the image outside the channel walls were masked out. Next, the swimmer was searched using inbuilt functions in MATLAB based on Circular Hough Transform. Then, a region of interest (ROI), based on the direction of motion of swimmer, was defined corresponding to the particle raft in-front of the swimmer. The start of the ROI was considered at a fixed azimuthal distance of $1.5 r_{\textrm{s}}$ from the swimmer center. The end of the ROI was defined where the speed of the particles drops (Fig.\ref{figure_raft_properties}b), for short rafts, or till the end of the channel in the image, for long rafts. Before counting the fraction covered by particles in the ROI, the image was binarized using a fixed intensity threshold. Finally, the ratio of the black pixels, corresponding to particles, to the total pixels within the region of interest was calculated as the average packing fraction of the raft.

\subsection{Estimation of drag on the raft from experiments} \label{Appendix F_d_raft expt}\
The drag on the particle raft was estimated from the experiments using the mean speeds of the swimmer in presence of particles $v_{\textrm{s}}(\phi_{\textrm{ini}})$ and in absence of particles $v_{\textrm{s}}(\phi_{\textrm{ini}}=0)$. Equations \ref{force balance eqn} and \ref{force_balance_with_particles} were considered for force balance on the swimmer under steady state motion in absence and in presence of particles, respectively. Substituting eqn.\ref{force balance eqn} into eqn.\ref{force_balance_with_particles} yielded the experimental estimate of the drag on the raft  
\begin{equation}
    F^{\textrm{expt}}_{\textrm{d,raft}} = \frac{16}{3}\mu r_{\textrm{s}} \left [ \frac{v^{3/2}_{s}(\phi_{\textrm{ini}}=0)-v^{3/2}_{s}(\phi_{\textrm{ini}})}{v^{1/2}_{s}(\phi_{\textrm{ini}})} \right].
\end{equation}

\noindent\textbf{Data accessibility.} Data are available at https://osf.io/6u2dt/. 


\noindent\textbf{Competing interests.} There are no conflicts to declare.

\noindent\textbf{Funding.} This work was supported by the NASA (80NSSC21K0287) and the NSF CBET-2401507. 

\noindent\textbf{Acknowledgments.} The authors would like to thank J. Rotheram and T.J. Sheppard for their help and initial contributions in the present work. We also thank Y. Fu, Dr. Z. Liu, C. Fowler and Dr. J. Yuk for useful discussions.

\bibliography{swimmer}
\end{document}